\documentclass[epj]{svjour}
\usepackage{graphics}
\begin{document}
\title{Shell-model structure of exotic nuclei beyond $^{132}$Sn}
\author{A. Covello\thanks{covello@na.infn.it} \and L. Coraggio \and A. Gargano \and N.Itaco}
\institute{Dipartimento di Scienze Fisiche, Universit\`a di Napoli Federico II, and
Istituto Nazionale di Fisica Nucleare, \\ Complesso Universitario di Monte S. Angelo, Via Cintia, I-80126 Napoli, Italy}

\date{Received: date / Revised version: date}
%
\abstract{ We report on a study of exotic nuclei around doubly magic $^{132}$Sn in terms of the shell model employing a realistic effective interaction derived from the CD-Bonn nucleon-nucleon potential. The short-range repulsion of the bare potential is renormalized by constructing a smooth low-momentum potential, $V_{\rm low-k}$, that is used directly as input for the calculation of the effective interaction. In this paper we focus attention on the nuclei 
$^{134}$Sn and $^{135}$Sb which, with an N/Z ratio of 1.68 and 1.65, respectively, are at present the most exotic nuclei beyond $^{132}$Sn for which information exists on excited states.
Comparison shows that the calculated results for both nuclei are in very good agreement with the experimental data. We present our predictions of the hitherto unknown spectrum of 
$^{136}$Sn. 
 }
\maketitle
\section{Introduction}
\label{intro}
The study of neutron-rich nuclei around doubly magic $^{132}$Sn is a subject of special interest, as it offers the opportunity to explore for possible changes in nuclear structure properties when moving toward the neutron drip line. In this context, great attention
is currently focused on exotic nuclei beyond the $N=82$ shell closure. This is motivated by the fact that some of the data that have become available appear to be at variance with what one might expect by extrapolating the existing results for $N < 82$  nuclei. In particular, some peculiar properties have been recently observed in the two nuclei $^{134}$Sn and $^{135}$Sb which, with an N/Z ratio of 1.68 and 1.65, respectively, are at present the most exotic nuclei beyond $^{132}$Sn for which information exists on excited states. This is the case of  the first $2^{+}$ state in $^{134}$Sn which, lying at 726 keV excitation energy, is the lowest first-excited $2^{+}$ level observed in a semi-magic even-even nucleus over the whole chart of nuclides. As for  $^{135}$Sb, there is a significant drop in the energy of the lowest-lying $5/2^{+}$ state as compared to the values observed for the Sb isotopes with $N \leq 82$. 

These data might be seen as the onset of a \linebreak shell-structure modification which, starting at $N=83-84$, is expected to produce more evident effects for larger neutron excess. From this viewpoint, the anomalously low position of the $5/2^{+}$ state in $^{135}$Sb may be attributed to a downshift of the $d_{5/2}$ proton level relative to the $g_{7/2}$ one caused by a more diffuse nuclear surface produced by the two neutrons beyond the 82 shell closure. Actually, a good description of $^{135}$Sb has been obtained in the shell-model study of refs. \cite{Shergur02,Shergur05a} using experimental single-particle (SP) energies with this downshift set at 300 keV. However, while the same kind of calculation also provides a satisfactory agreement with experiment for $^{134}$Sn \cite{Shergur02}, this is not the case for the one proton, one neutron nucleus $^{134}$Sb \cite{Shergur05b}. 

In recent work  \cite{Coraggio05,Coraggio06,Covello05a} we have shown that the properties of these three nuclei are well accounted for by a unique shell-model Hamiltonian with SP energies taken from experiment and the two-body effective interaction derived from the CD-Bonn nucleon-nucleon ($NN$) potential \cite{Machleidt01}. The short-range repulsion of the latter has been renormalized by use of the low-momentum potential $V_{\rm low-k}$ \cite{Bogner02}, as we shall briefly discuss in sect. \ref{sec:1}.

In this paper, we focus attention on the two nuclei $^{134}$Sn and $^{135}$Sb. Based on the good agreement between our results and the available experimental data obtained in 
\cite{Coraggio05,Covello05a}, we find it interesting to report here the complete calculated low-energy spectra of these two nuclei. We hope that this may stimulate, and be helpful to,  future experiments on these highly important nuclei. With the same motivation, we also present our predictions of the hitherto unknown spectrum of $^{136}$Sn with four neutrons outside $^{132}$Sn, which makes an N/Z ratio of 1.72.

The outline of the paper is as follows. In sect. \ref{sec:1} we give a brief description of the theoretical framework in which our realistic shell-model calculations have been performed. In sect. \ref{sec:2} we give some details of the calculations and present our results together with the experimental data available for $^{134}$Sn and $^{135}$Sb. Some concluding remarks are given in sect. \ref{sec:3}.

\section{Outline of theoretical framework}
\label{sec:1}

The starting point of any realistic shell-model calculation is the free $NN$ potential. There are, however, several high-quality potentials, such as Nijmegen I and Nijmegen II  \cite{Stoks94}, Argonne $V_{18}$ \cite{Wiringa95}, and CD-Bonn \cite{Machleidt01}, which fit equally well ($\chi^2$/datum $\approx 1$) the $NN$ scattering data up to the inelastic threshold. This means that their on-shell properties are essentially identical, namely they are phase-shift equivalent.   
In our shell-model calculations we have derived the effective interaction from the CD-Bonn  potential. This may raise the question of how much our results may depend on this choice of the $NN$ potential. We shall comment on this point later in connection with the $V_{\rm low-k}$ approach to the renormalization of the bare $NN$ potential.

The shell-model effective interaction $V_{\rm eff}$ is defined, as usual, in the following way. In principle, one should solve a nuclear many-body Schr\"odinger equation of the form 
\begin{equation}
H\Psi_i=E_i\Psi_i ,
\end{equation}
with $H=T+V_{NN}$, where $T$ denotes the kinetic energy. This full-space many-body problem is reduced to a smaller model-space problem of the form
\vspace{-.1cm}
\begin{equation}
PH_{\rm eff}P \Psi_i= P(H_{0}+V_{\rm eff})P \Psi_i=E_iP \Psi_i .
\end{equation}
\noindent Here $H_0=T+U$ is the unperturbed Hamiltonian, $U$ being an auxiliary potential introduced to define a convenient single-particle basis, and $P$ denotes the projection operator onto the chosen model space.

As pointed out in the Introduction, we ``smooth out" the strong
repulsive core contained in the bare $NN$ potential $V_{NN}$ by
constructing a low-momentum  potential $V_{\rm low-k}$. 
This is achieved by integrating out the \linebreak high-momentum modes
of $V_{NN}$ down to a cutoff momentum  $\Lambda$. 
This integration is carried out with the requirement that the deuteron
binding energy and phase shifts of $V_{NN}$ up to $\Lambda$ are preserved by $V_{\rm low-k}$. 
A detailed description of the derivation of $V_{\rm low-k}$ from $V_{NN}$ as well as a discussion of its main features can be found in refs. \cite{Bogner02,Covello05b}. However, we should mention here that shell-model effective interactions derived from different phase-shift equivalent $NN$ potentials through the $V_{\rm low-k}$ approach lead to very similar results
\cite{Covello05b}. In other words, $V_{\rm low-k}$ gives an approximately unique representation of the $NN$ potential.

Once the $V_{\rm low-k}$ is obtained, we use it as input interaction  for the calculation of the matrix elements of the shell-model effective interaction. The latter is derived by employing a folded-diagram method, which was previously applied to many nuclei using $G$-matrix interactions \cite{Covello01}. Since $V_{\rm low-k}$ is already a smooth potential, it is no longer necessary to calculate the $G$ matrix. We therefore perform shell-model calculations following the same procedure as described, for instance, in 
\cite{Jiang92,Covello97}, except that the $G$ matrix used there is replaced by $V_{\rm low-k}$. More precisely, we first calculate the so-called $\hat{Q}$-box \cite{Kuo90} including diagrams up to second order in the two-body interaction. The shell-model effective interaction is then obtained by summing up the $\hat{Q}$-box folded diagram series using the Lee-Suzuki iteration method \cite{Suzuki80}.

\section{Calculations and results}
\label{sec:2}

In our calculations we assume that $^{132}$Sn is a closed core and let the valence
neutrons occupy the six levels $0h_{9/2}$, $1f_{7/2}$, $1f_{5/2}$, $2p_{3/2}$,
$2p_{1/2}$, and  $0i_{13/2}$ of the 82-126 shell, while for the odd proton in $^{135}$Sb
the model space includes the five  levels  $0g_{7/2}$, $1d_{5/2}$, $1d_{3/2}$, $2s_{1/2}$,
and $0h_{11/2}$ of the 50-82 shell. 
As mentioned in the previous section, the two-body matrix elements of the effective interaction are derived from the CD-Bonn $NN$ potential renormalized through the 
$V_{\rm low-k}$ procedure with a cutoff momentum $\Lambda=2.2$ fm$^{-1}$. This value of
$\Lambda$ is in accord with the criterion given in ref. \cite{Bogner02}. The computation of the diagrams included in the $\hat{Q}$-box is performed within the harmonic-oscillator basis using  intermediate states composed of all possible hole states and particle states restricted  to the five shells above the Fermi surface. The oscillator parameter used is $\hbar \omega = 7.88$ MeV.

As regards the SP energies, they have been taken from experiment. In particular, the spectra  \cite{nndc} of $^{133}$Sb and $^{133}$Sn have been used to fix the proton
and neutron SP energies, respectively. The only exceptions are the proton $\epsilon_{s_{1/2}}$ and neutron $\epsilon_{i_{13/2}}$, whose corresponding levels are still missing. Their values have been taken from refs. \cite{Andreozzi97} and 
\cite{Coraggio02}, respectively, where it is discussed how they are determined.
All the adopted values are reported in ref. \cite{Coraggio05}.

\begin{figure}[h]
\resizebox{0.48\textwidth}{!} {%
\includegraphics{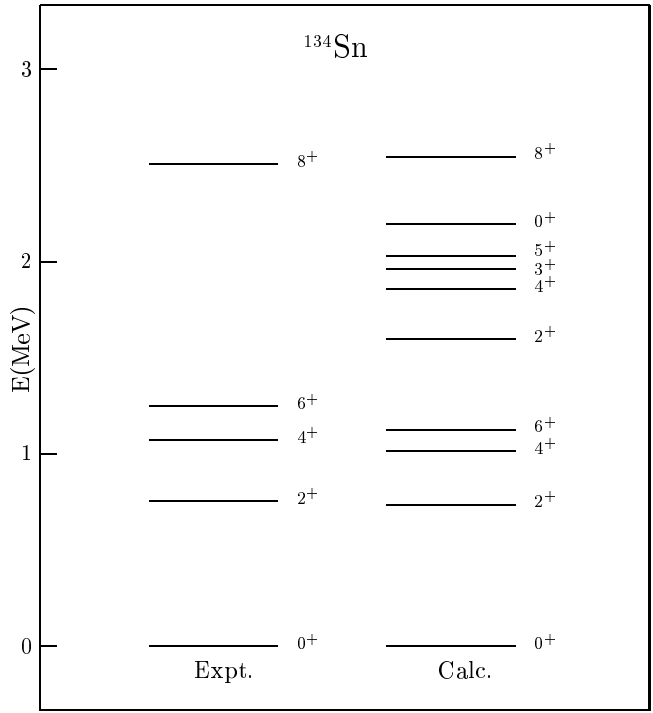} 
}
\caption{Experimental and calculated spectrum of $^{134}$Sn}
\label{fig:1}       
\end{figure}
Let us start with $^{134}$Sn. From fig. 1 we see that while the theory reproduces very well all the observed levels \cite{Zhang97,Korgul00}, it also predicts, in between the $6^+$ and $8^+$ states, the existence of five states with spin $\leq 5$.  Clearly, the latter could not be seen from the $\gamma$-decay of the $8^+$ state populated in the spontaneous fission experiment of ref. \cite{Korgul00}.

Very recently, the $B(E2;0^+ \rightarrow 2_1^+)$ value in $^{134}$Sn has been measured \cite{Beene04} using Coulomb excitation of neutron-rich radioactive ion beams. We have calculated this $B(E2)$ with an effective neutron charge of $0.70\,e$, according to our early study \cite{Coraggio02}. We obtain $B(E2;0^+ \rightarrow 2_1^+)$ = 0.033 $e^2$b$^2$,
in excellent agreement with the experimental value 0.029(4) $e^2$b$^2$.

\begin{table}[h]
\caption{Experimental and calculated excitation energies (in MeV)  for $^{135}$Sb}
\label{tab:1}       
\begin{tabular}{rlrl}
\hline\noalign{\smallskip}
\multicolumn {2} {c} {Calc.} & \multicolumn {2} {c} {Expt.}  \\
$J^{\pi}$ & E & $J^{\pi}$ & E  \\
\noalign{\smallskip}\hline\noalign{\smallskip}
$\frac{7}{2}^{+}$ & 0.0 & $\frac{7}{2}^{+}$ & 0.0 \\
$\frac{5}{2}^{+}$ & 0.391 & $\frac{5}{2}^{+}$ & 0.282 \\
$\frac{3}{2}^{+}$ & 0.509 & $\frac{3}{2}^{+}$ & 0.440 \\
$\frac{1}{2}^{+}$ & 0.678 & & \\
$\frac{11}{2}^{+}$ & 0.750 & $\frac{11}{2}^{+}$ & 0.707 \\
$\frac{9}{2}^{+}$ & 0.813 & $\frac{9}{2}^{+}$ & 0.798 \\
$\frac{5}{2}^{+}$ & 0.924 & &  \\
$\frac{7}{2}^{+}$ & 0.938 & & \\
&  & $\frac{7}{2}^{+}$ & 1.014 \\
$\frac{7}{2}^{+}$ & 1.031 & & \\
$\frac{9}{2}^{+}$ & 1.108 & $\frac{9}{2}^{+}$ & 1.027 \\
$\frac{15}{2}^{+}$ & 1.124 & $\frac{15}{2}^{+}$ & 1.117 \\
$\frac{9}{2}^{+}$ & 1.144 & &  \\
$\frac{5}{2}^{+}$ & 1.146 & &  \\
& & $\frac{5}{2}^{+}$ & 1.113 \\
$\frac{5}{2}^{+}$ & 1.182 & &  \\
$\frac{13}{2}^{+}$ & 1.209 & &  \\
$\frac{7}{2}^{+}$ & 1.231 & $\frac{7}{2}^{+}$ & 1.207 \\
$\frac{3}{2}^{+}$ & 1.263 & & \\
$\frac{19}{2}^{+}$ & 1.268 & $\frac{19}{2}^{+}$ & 1.342 \\
\noalign{\smallskip}\hline
\end{tabular}
\end{table}

The calculated and observed levels \cite{Shergur05b,nndc} of $^{135}$Sb up to about 1.3 MeV excitation energy are reported in table 1. We see that the agreement between theory and experiment is very good. It is worth noting that our calculation reproduces the observed $5/2^{+}$ state within 100 keV. This is a relevant result as it argues against the suggestion \cite{Shergur02} that the low position of this state, assumed  to be essentially of single-particle character, is related to a relative shift of the proton $d_{5/2}$ and $g_{7/2}$ orbits induced by the neutron excess.
From table 1 it can be seen that we predict several low-lying excited states which have no experimental counterpart. We hope that these predictions will be verified by further experimental work.

In the very recent work of refs. \cite{Mach05,Korgul} the lifetime of the $5/2^{+}$
state in $^{135}$Sb has been measured. A very small upper limit for the $B(M1)$, $0.29\cdot10^{-3} \, \mu_{N}^{2}$, was found, thus evidencing a strongly hindered transition, which may be seen \cite{Shergur02} as a confirmation of the single-particle nature of the $5/2^{+}$ state. This is at variance with the outcome of our calculation. 
In fact, as a consequence of the neutron-proton interaction, we have found that the $5/2^{+}$ state is of admixed nature. We have calculated the $B(M1;5/2^{+} \rightarrow 7/2^{+})$  making use of an effective $M1$ operator which includes first-order diagrams in $V_{\rm low-k}$. Our predicted value is 
$4.0\cdot10^{-3} \, \mu_{N}^{2}$. Keeping in mind that in our calculation we do not include any meson-exchange correction, the agreement between the experimental and calculated $B(M1)$ may be considered quite satisfactory. 

\begin{table}[h]
\caption{Calculated excitation energies (in MeV) for $^{136}$Sn}
\label{tab:1}       
\begin{tabular}{ll}
\hline\noalign{\smallskip}
$J^{\pi}$ & Calc. \\
\noalign{\smallskip}\hline\noalign{\smallskip}
$0^{+}$ & 0.0  \\
$2^{+}$ & 0.715 \\
$4^{+}$ & 1.020 \\
$6^{+}$ & 1.163   \\
$4^{+}$ & 1.170 \\
$2^{+}$ & 1.329 \\
$2^{+}$ & 1.417 \\
$5^{+}$ & 1.465 \\
$4^{+}$ & 1.660   \\
$3^{+}$ & 1.724  \\
$8^{+}$ & 1.811 \\
\noalign{\smallskip}\hline
\end{tabular}
\end{table}

In table 2 we report the calculated excitation energies of $^{136}$Sn  up to about 1.8 MeV.
We find that three lowest states, $2^+, 4^+$, and $6^+$, lie at practically the same energy as in $^{134}$Sn. Above the $6^+$ level there are seven states in an energy interval of about 650 keV. This pattern is quite different from that predicted for the spectrum of $^{134}$Sn, where a rather pronounced gap (about 0.5 MeV wide) exists between the $6^+$ state and the next excited state with $J^\pi=2^+$.

\section{Summary}
\label{sec:3}
We have presented here the results of a shell-model study of neutron-rich nuclei around 
$^{132}$Sn, focusing attention on  $^{134}$Sn and $^{135}$Sb which are at present the most exotic nuclei with valence neutrons beyond $N=82$. The two-body effective interaction has been derived by means of a $\hat Q$-box folded-diagrams method from the CD-Bonn $NN$ potential, the short-range repulsion of the latter being renormalized by use of the low-momentum potential $V_{\rm low-k}$. 

Our results for both nuclei are in very good agreement with the observed spectroscopic properties. These results, considered along with  those obtained for $^{134}$Sb \cite{Coraggio06}, show that to explain the presently available data on neutron-rich nuclei beyond $^{132}$Sn there is no need to invoke shell-structure modifications. These data, however, are still rather scanty and we have found it challenging to make predictions which may stimulate further experimental efforts to study this kind of exotic nuclei. In particular, we hope that our predictions for the  hitherto unknown $^{136}$Sn will be verified in a not too distant future.

\begin{acknowledgement}
This work was supported in part by the Italian Ministero \linebreak dell'Istruzione, dell'Universit\`a e della Ricerca  (MIUR).
\end{acknowledgement}

\end{document}